\documentclass{kluwer}    

\newdisplay{guess}{Conjecture}

\begin{document}                                                                                   
\begin{article}
\begin{opening}         
\title{Cosmological perturbations of an expanding brane in an anti-de Sitter bulk : a short review} 
\author{Nathalie \surname{Deruelle}}  
\runningauthor{Nathalie Deruelle}
\runningtitle{Brane cosmological perturbations}
\institute{Institut d'Astrophysique de Paris,
GReCO, FRE 2435 du CNRS,
 98 bis boulevard  Arago, 75014, Paris, France}
\date{October 22nd 2002}

\begin{abstract}
Since the Randall-Sundrum 1999 papers, braneworlds have been a favourite playground to test string inspired cosmological models. The subject has developped
into two main directions : elaborating more complex models in order to strenghten the connection with string theories, and trying to confront them with observations, in
particular the Cosmic Microwave Background anisotropies. We review here the latter and see that, even in the simple, ``paradigmatic", case of a single expanding brane in a 5D
anti-de Sitter bulk, there is still a missing link between the ``view from the brane" and the ``view from the bulk" which prevents definite predictions.
\end{abstract}

\end{opening}           

\section {Introduction}  

Since the now classic 1999 papers by Randall and Sundrum [2], there has been a growing interest for gravity theories in spacetimes with large extra dimensions and the idea
that our universe may be a four dimensional singular hypersurface, or ``brane", in a five dimensional spacetime, or ``bulk". 

The (second) Randall-Sundrum scenario, where our  universe is represented by a four dimensional quasi-Minkowskian edge of a double-sided perturbed anti-de Sitter
spacetime, or ``$Z_2$-symmetric" bulk, was the first model where the linearized Einstein equations were found to hold on the brane, apart from small $1/r^2$
corrections to Newton's potential [2] [5]. 

 Cosmological models were then soon to be built, where the brane, instead of flat, is taken to be a Robertson-Walker spacetime, and it was shown that such ``braneworlds" can
tend at late times to the standard Big-Bang model and hence represent the observed universe [4].

The subject has since developped into two main directions~:

On one hand, more complex models were elaborated, in order to turn the Randall-Sundrum scenario from a ``toy" to  a more ``realistic" low energy limit of string theories.
That included considering two brane models, studying the dynamics and stabilisation of the distance between the branes (the ``radion"), allowing for colliding
branes, as well as turning the 5D cosmological constant into a scalar field living in the bulk, correcting Einstein's equations with a Gauss-Bonnet term, etc.

On another hand, effort has been devoted to try and confront these models  with observations, in particular the Cosmic Microwave Background anisotropies. In order to do
so, various set ups to study the perturbations of braneworlds have been proposed and compared to the perturbations of standard, four dimensional,
Friedmann universes. 

We concentrate here on the simple, ``paradigmatic", case of a single expanding brane in a 5D anti-de Sitter bulk and briefly review the 40 odd papers dealing with the
cosmological perturbations of this toy model. As we shall see, they all have up to now stalled on the problem of solving, in a general manner,  the Israel ``junction conditions"
(that is the Einstein equations integrated across the brane) which relate the matter perturbations on the brane and the perturbations in the bulk.

\section{The brane gravity equations}

 A first approach to obtain the equations which govern gravity on the brane, called for short the ``view from the brane",  is to project the bulk 5D Einstein equations, ${\cal
G}_{AB}=\Lambda\gamma_{AB}$, on the brane. To do so it is convenient to (1) introduce a gaussian normal coordinate system where the brane is located at $y=0$, (2) 
expand the metric in   Taylor series in $y$, (3) write the Einstein 5D equations at lowest order in $y$, (4) relate, by means of the Israel junction conditions, the first order term  of the
Taylor expansion of the metric (that is, the extrinsic curvature of the brane) to the brane tension and stress-energy tensor,  and, (5),  get the Shiromizu-Maeda-Sasaki (SMS) equations for
gravity on the brane [3] 
$$G_{\mu\nu}=8\pi GT_{\mu\nu}+{(8\pi G)^2\over\Lambda}S_{\mu\nu}+E_{\mu\nu}$$
$$D_\mu T^\mu_\nu=0$$
$$-R=8\pi G T+{(8\pi G)^2\over\Lambda}S\quad\Longleftrightarrow\quad E=0$$
where $G$ is Newton's constant, where $G_{\mu\nu}$ and $R$ are the brane Einstein tensor and Ricci scalar, where $S_{\mu\nu}$ is some tensor quadratic in $T_{\mu\nu}$, and
where the  projected Weyl tensor
$E_{\mu\nu}$ is related to the second order term of the Taylor expansion of the metric (see also [32]). 

If we impose the brane to be a (flat) Robertson-Walker type universe with scale factor $a$ and Hubble parameter ${\dot a\over a}\equiv H$, then the second equation is the
conservation equation~: 
$$\dot\rho+3H(\rho+p)=0$$ $\rho$ and $p$ being the energy density and pressure of the brane cosmological fluid~; the third equation gives the modified
Friedmann (BDEL) equation [4]~: 
$$H^2={8\pi G\over 3}\rho\left(1+{4\pi G\rho\over\Lambda}\right)+{c\over a^4}$$ and the first gives $E_{\mu\nu}$ (which is zero if $c=0$), that is the
metric off the brane up to second order in $y$. By iteration one gets the (BDL) metric everywhere in the bulk [1]. It looks complicated but one soon realizes [8] that (for $c=0$)
the bulk is nothing but 5D anti-de Sitter spacetime (see also [11]).

The ``view from the bulk", on the other hand, consists in considering a 5D Einstein manifold (such that ${\cal G}_{AB}=\Lambda\gamma_{AB}$) and
imposing a foliation by maximally symmetric 3-spaces. One then immediatly gets, from Birkhoff's theorem, that the 5D manifold is 5D anti-de Sitter spacetime
(Schwarzschild-AdS5 if
$c\neq0$). In coordinates adapted to the symmetries of the bulk (e.g. conformally minkowskian if the brane is spatially flat), the bulk metric looks simple
($ds^2={6\over\Lambda( X^4)^2}\eta_{AB}dX^AdX^B$), but the equation for the brane is (slightly) more complicated than in gaussian normal coordinates
($X^4=\sqrt{6\over\Lambda}{1\over a}\equiv A$, $X^0=\sqrt{1+A'^2}$), see [6] and e.g. [25].

\section{Braneworld perturbations~: the view from the brane}

In this approach, one concentrates on  perturbing the  SMS equations around the BDEL, Friedmann-modified, background brane solution (with $c=0$).

A first step is to assume $E_{\mu\nu}=0$ at linear order. The SMS equations then differ from the standard Einstein equations only by the presence of the $S_{\mu\nu}$ term.
Standard perturbation theory (either in a ``covariant" [30] or ``gauge invariant" formulation) can then be applied. One result one can reach is, for example, that the
conditions for inflation on the brane are different from the standard 4D case because of the presence of $S_{\mu\nu}$ with, as a consequence, that the initial density spectrum
is enhanced, and the initial gravitational spectrum less so, as compared to the standard 4D case. See e.g., ref [9] [21] [31] [38] [46].

A second step consists in isolating in $E_{\mu\nu}$ the bits which prevent the SMS equations to close on the brane. It turns out that it is its transverse traceless part, $
P_{\mu\nu}$. However when one considers scalar perturbations only, then $P_{\mu\nu}=D_\mu D_\nu P$, and this term drops out of the SMS equations on super
Hubble scales. The system is then closed on large scales  but, instead of involving only one master variable (the perturbation of the inflaton basically), it  also involves $E_{00}$,
which acts as a kind of second scalar field and induces, on top of the standard inflationary adiabatic perturbations, isocurvature ones [14] [22] [30].

In [26], the authors recovered the previous results using a gauge invariant, rather than covariant, formalism. They also showed that, if the SMS equations for the initial
density perturbation spectrum indeed closes on the brane, the equations governing the $g_{00}$ perturbation were, on the other hand, not closed, so that the Sachs-Wolfe
contribution to the CMB anisotropies could not be predicted without further knowledge of the bulk.

One can nevertheless write a Boltzmann code including the contributions of $S_{\mu\nu}$ and $E_{\mu\nu}$ [39]. If one then assumes some specific behaviour for
$E_{\mu\nu}$ [37], the CMB anisotropies can be calculated, see [45] for preliminary results.

\section{The view from the bulk in coordinates adapted to the brane}
As we have already mentionned, the anti-de Sitter metric looks complicated when written in gaussian coordinates in which the equation for the brane is $y=0$. One can however
perturb this BDL metric [15] and write the perturbation of the 5D Einstein tensor in terms of the perturbation of the 4D Einstein tensor plus terms involving first an second
$y$-derivatives of the metric perturbations. The first derivative terms are expressed in terms of the perturbations of the stress-energy tensor of matter on the
brane thanks to the Israel junction conditions. As for the second order derivatives they remain undetermined and can be interpreted as some kind of extra ``seeds" in the
4D perturbation equations [23]. Of course these final equations must be equivalent, and were shown to be equivalent [23],  to the perturbed SMS equations, with the second order
derivative identified with the projected Weyl tensor $E_{\mu\nu}$. 

A drawback of this choice of gauge is, first, that the perturbation equations are very complicated, so that the regularity conditions which one must impose on the bulk
perturbations at the AdS5 horizon have not yet been implemented; second, the brane bending effect is put under the rug [25] [34], which renders the boundary conditions even more
difficult to implement.

However some partial results could be obtained [19] [41]. For example, if the brane is 4D de Sitter spacetime, then the equations for the tensor perturbations $T$ can be
integrated by separation of the $t$ and $y$ variables (see [36] for an explanation of this simplification). They can be written as an infinite tower of modes ($T=\int
dm\,\phi_m(t){\cal E}_m(y)$), and the normalisation of the ${\cal E}_m(y)$ part imposed by the bulk boundary conditions [2] [5], first suppresses the non zero modes, and, second,
yields a modified normalisation of $\phi_0(t)$ when one quantizes it. Hence a spectrum of gravitational waves different from the standard 4D one [19]. The same procedure was
applied to the vectorial modes [24], with the somewhat surprising result that they can be normalized only if some matter vorticity is present.

The 5D ``longitudinal gauge" adopted by a number of authors [16] [27] [28] is the closest to the very commonly used 4D longitudinal gauge, which allows to write the
perturbation equations under a fairly familiar form. However, since this gauge is completely fixed, the brane cannot a priori be placed at $y=0$~;  it can  be placed at $y=0$ only
if there are no matter anisotropic stresses. To include brane bending and anisotropic stresses it is therefore necessary to go to another gauge if one wants to keep the brane at
$y=0$, which spoils a bit the form of the perturbation equations. Of course the perturbation equations in the gaussian normal and the longitudinal gauge must be equivalent, and
were shown to be [34].

In order to ease the passage 
from one gauge to another, various fully gauge invariant formalisms were proposed [12] [13] [20] [29] [33] [40], which have the advantage of
expressing the scalar, vector and tensor perturbations in terms of 3 independent master variables whose evolution equations are known. Connection with the gaussian normal and
the longitudinal gauge was performed in [29] [34]  [42].

\section{The view from the bulk in coordinates adapted to the bulk}

As we have mentionned earlier,  the AdS5 spacetime metric is very simple when written in conformally minkowskian coordinates. The perturbations equations are also very
simple in this coordinate system and can be explicitely solved in, e.g., the standard tansverse-traceless gauge, see [18] [25] [35] [43], or in a gauge invariant way [33]. In that
background coordinate system, the regularity conditions on the graviton modes at the AdS5 horizon can also easily be discussed, see e.g. [7] [18] [25] ; for example one may only
keep the outgoing modes. Finally the brane bending degree of freedom is also easily taken into account by perturbing the position of the brane.

It is then straightforward to write the extrinsic curvature of this bent brane in a perturbed AdS5 spacetime and relate it, by means of the Israel junction conditions, to the
perturbations of the stress-energy tensor of the matter inside the brane [25]. In this approach then, the perturbations of the brane stress-energy tensor are determined in terms
of the brane bending and the (regular) 5D graviton modes. They can then be split into perturbations of the brane matter fields on one hand, and ``seeds" on the other. If one assumes
or imposes the absence of seeds, this approach gives straightforwardly the allowed brane bending and 5D gravitons compatible with this constraint. Particular cases
have been studied, for example the case of an inflating brane [25].  The serious drawback of this ``view from he bulk" approach is that the connection with the previous ones is not
straightforward, has not been done yet and hence has not yet given the missing piece of information, that is the expression of the projected Weyl tensor $E_{\mu\nu}$ which is
needed in order to implement the existing Boltzmann codes to yield the CMB anisotropies.

There exists however a more promising approach [18] [35], which consists in starting, as above,  with the brane bending and the (regular) 5D graviton modes expressed  as
perturbations of AdS5 spacetime in conformally minkowskian coordinates, and then in performing the (``large") coordinate transformation which brings the conformally
minkowskian coordinates to the gaussian normal (BDL) ones before projecting them on the brane. However, here too,  the connection with the ``view from the brane" approaches
has not been completed yet, although partial results have already been reached, for example the fact that the isocurvature brane mode found in [22] would correspond to a
divergent bulk mode [35] [43].

\section{Conclusion}

To summarize the situation in one paragraph~: when one treats the braneworld cosmological perturbations in a strict brane point of view, one stumbles on the problem of
finding the expression for the projected Weyl tensor, and that alone prevents from predicting the CMB anisotropies. When one attempts to treat the problem by looking at the bulk
perturbations in coordinate systems adapted to the brane, say, gaussian normal, then one obtains discouragingly complicated perturbation equations that can be solved only in the
very particular dS4 brane case. Finally, when one adopts a coordinate system adapted to the bulk geometry, then the brane perturbations are obtained under a form which is very
different from the familiar, 4D perturbations equations, so that standard  Boltzmann codes cannot be used to yield the CMB anisotropies.

The problem of computing the CMB anisotropies generated by brane\-worlds will be solved when the gap between the two approaches is bridged, along the lines of [42] [43] or [44].
In order to do so, it may prove useful to study toy models, such as the case of a de Sitter brane, or induced gravity models in which the bulk is nothing but 5D Minkowski
spacetime.

\end{article}
\end{document}